\documentclass[aps,prl,twocolumn,showpacs,floatfix,superscriptaddress]{revtex4-1}
\usepackage{amsmath,amssymb,graphicx,color}
\usepackage{bm}
\bmdefine{\bfDelta}{\Delta}
\bmdefine{\bfsigma}{\sigma}
\usepackage[colorlinks,citecolor=red]{hyperref}
\newcommand\unit[1]{\operatorname{#1}}

\newcommand\ket[1]{\mathinner{\lvert{\textstyle#1}\rangle}}
\newcommand\braket[1]{\mathinner{\langle{\textstyle#1}\rangle}}

\let\up=\uparrow
\let\down=\downarrow

\begin{document}

\title{Negative Tunneling Magneto-Resistance in Quantum Wires with Strong Spin-Orbit Coupling}
\author{Seungju Han}
\affiliation{Department of Physics, Korea University, Seoul 136-701, Korea}

\author{Lloren\c{c} Serra}
\affiliation{IFISC (CSIC-UIB) and Department of Physics, 
University of the Balearic Islands
E-07122 Palma de Mallorca, Spain}

\author{Mahn-Soo Choi}
\affiliation{Department of Physics, Korea University, Seoul 136-701, Korea}

\begin{abstract}
We consider a two-dimensional magnetic tunnel junction of
the FM/I/QW(FM+SO)/I/N structure, where FM, I and QW(FM+SO) stand for a
ferromagnet, an insulator and a quantum wire (QW) with both magnetic ordering
and Rashba spin-orbit (SOC), respectively.
The tunneling magneto-resistance (TMR) exhibits strong anisotropy and switches sign as the polarization direction varies relative to the QW axis, due to interplay among the one-dimensionality, the magnetic ordering, and the strong SOC of the QW.
The results may provide a possible explanation for the sign-switching anisotropic TMR recently observed in the LaAlO$_3$/SrTiO$_3$ interface.
\end{abstract}
\maketitle


The magnetic tunneling junction (MTJ) consisting of two ferromagnetic electrodes (FM) separated by a thin insulating barrier (I) is a prototype structure in the rapidly developing field of spintronics \cite{Wolf01a}. The tunneling magneto-resistance (TMR), depending on the relative magnetic polarization of the two FMs, is a
key issue not only for the spintronic applications but also for the study of fundamental magnetic properties \cite{Moodera10a,Miyazaki95a}.
Due to the spin selection rule the TMR, if any, is typically positive. Two exceptional cases have been known. One involves magnetic impurities in the tunnel barriers and is not surprising. The other (more important) case is associated with the resonant tunneling and spin-dependent interfacial phase shift in double-barrier FM/I/N/I/FM structures, where N represents a non-magnetic normal 
metal~\cite{Tsymbal03b,Sahoo05a,ChoiMS06d,ChoiMS06g,Yuasa02a}.

\begin{figure}
\centering
\includegraphics[width=30mm]{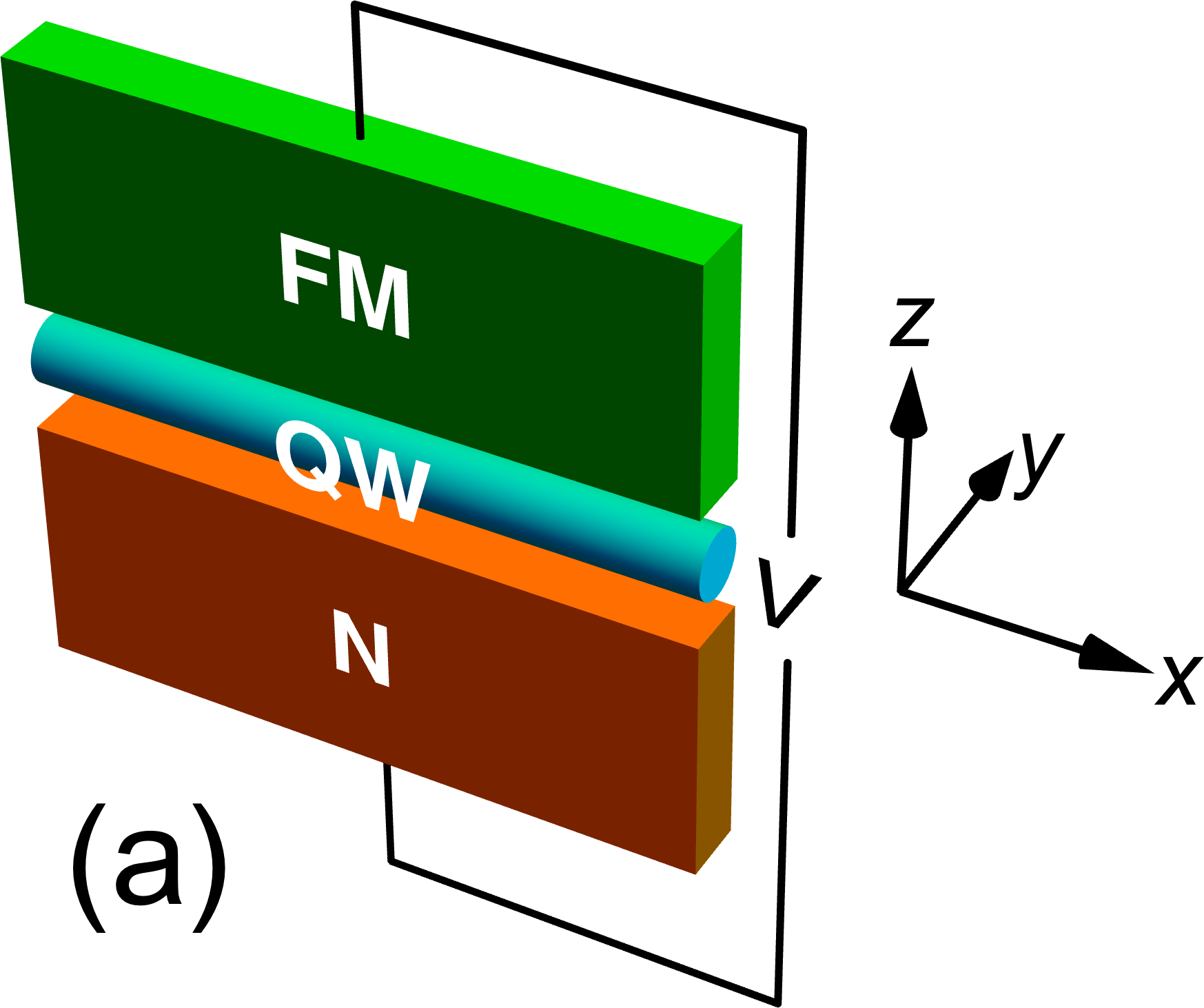}%
\includegraphics[width=25mm]{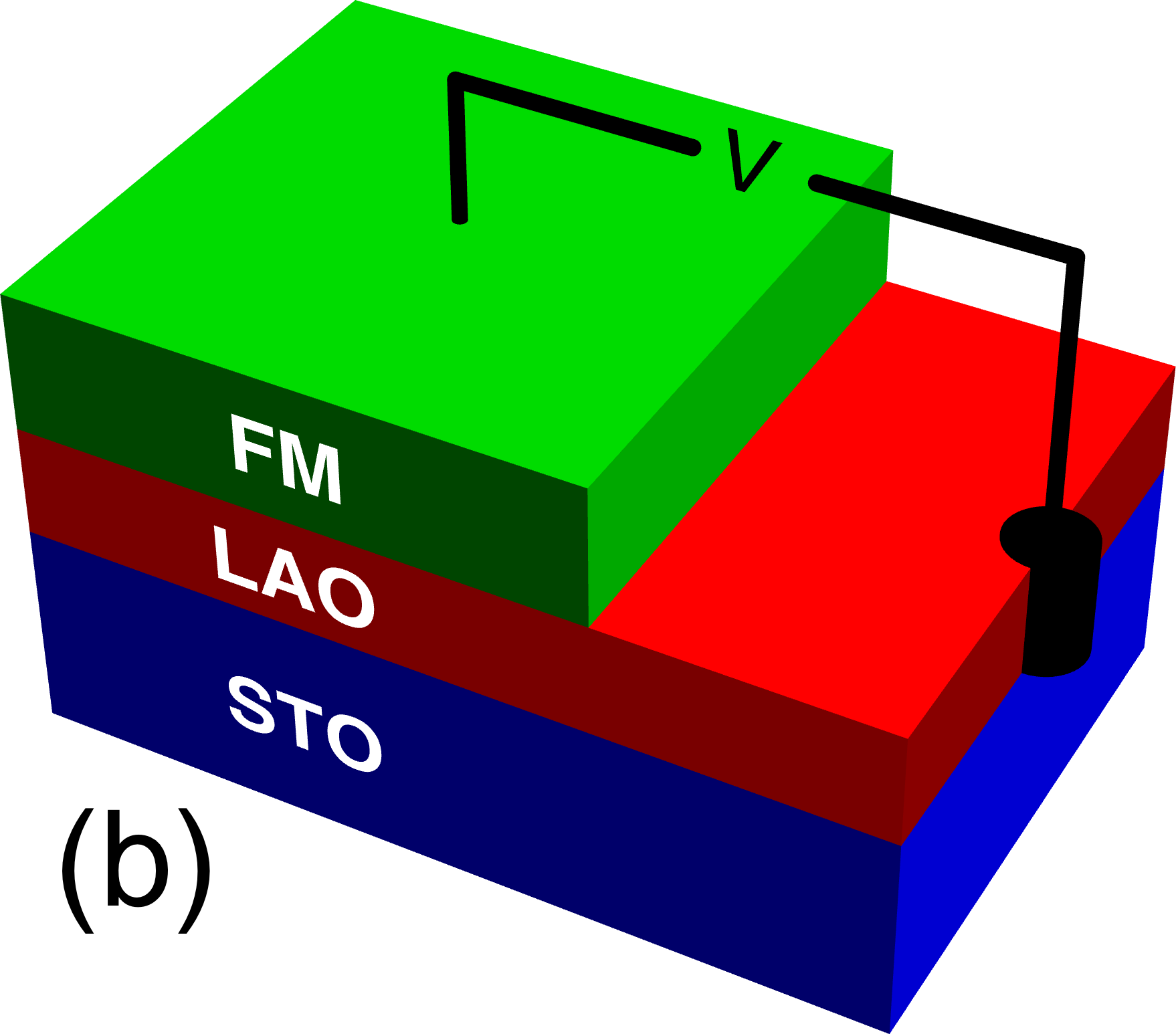}%
\includegraphics[width=25mm]{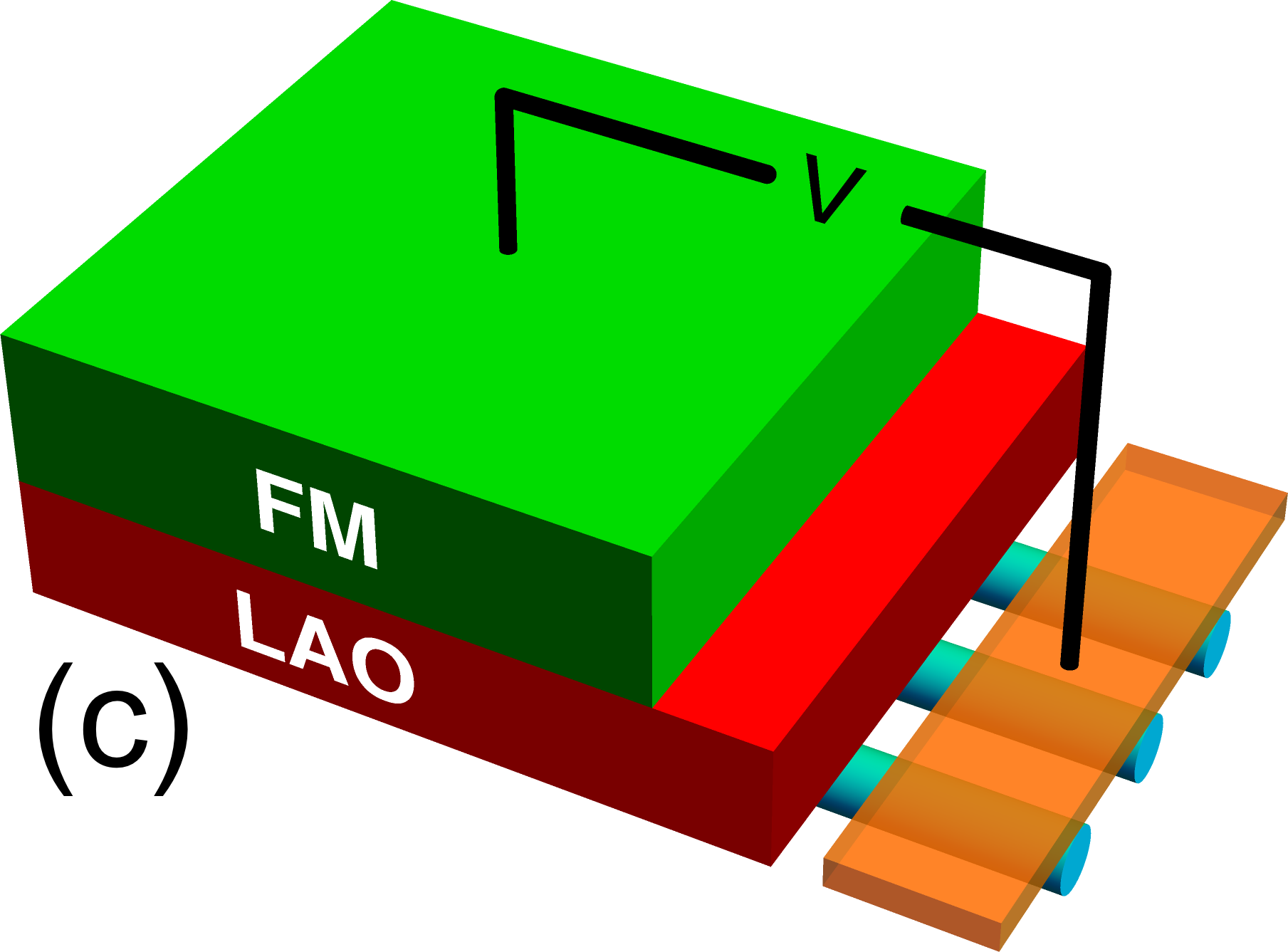}
\caption{(color online) (a) A double-barrier MTJ of the FM/I/QW(FM+SO)/I/FM structure. (b) A setup to measure the TMR between the top FM and the LAO/STO interface. (c) A simplified model of (b).}
\label{Paper::fig:1}
\end{figure}

In this work we explore another non-trivial example of negative TMR in a \emph{two-dimensional} (2D) double-barrier MTJ of the FM/I/QW(FM+SO)/I/N structure [see Fig.~\ref{Paper::fig:1} (a)], where QW(FM+SO) stands for a quantum wire (QW) with both magnetic ordering and Rashba spin-orbit coupling (SOC).
Our MTJ structure should be distinguished from more common 1D MTJs of the FM/I/QW/I/FM structure such as in \cite{Sahoo05a}, where the QW is non-magnetic and the junction interface is perpendicular to the axis of the QW. In our case, the QW itself has a magnetic ordering and the junction interface is parallel to its axis. 
Thus, transport occurs across, not along the QW.
We find that the TMR exhibits strong anisotropy and even changes sign as the polarization direction
of the FMs varies relative to the QW axis. This sign-switching anisotropic TMR is attributed to the interplay among the one-dimensionality, the magnetic ordering, and the strong SOC of the QW. It is interesting to recall that anisotropic TMR was previously studied in the FM/I/FM structure where the insulating barrier (not the  FMs) had SOC (see \cite{Matos-Abiague09a} and references therein), but the TMR remained positive without switching its sign.

Our MTJ structure is peculiar in that nanoscale QWs with both strong SOC and magnetic ordering are rare. However, an important motivation is the recent experiment \cite{Ngo14z} on the transition metal oxide interface between LaAlO$_3$ (LAO) and SrTiO$_3$ (STO) [see Fig.~\ref{Paper::fig:1} (b)], where the measured TMR is strongly anisotropic and switches sign as the magnetization direction varies in the interface plane. Since the LAO/STO interface was demonstrated a decade ago \cite{Ohtomo04a} to be metallic even though both LAO and STO are typical band insulators, it has attracted ever growing interest by exhibiting superconductivity \cite{Reyren07a}, ferromagnetism \cite{Brinkman07a} and even coexistence of both effects \cite{Li11a,Bert11a}. Despite a number of experimental studies of the system, the origins of magnetic ordering and superconductivity remain controversial \cite{Michaeli12a,Chen13a} and further studies are imperative.
The sign-switching anisotropic TMR \cite{Ngo14z} adds a fresh intriguing question concerning the magnetic properties of the LAO/STO interface. Our results below suggest one possible explanation for it in terms of our MTJ model mentioned above.

Indeed, a recent experiment \cite{Kalisky13a} suggests that the electric conduction in the LAO/STO interface [Fig.~\ref{Paper::fig:1} (b)] occurs mainly along the narrow paths associated with the twin boundaries in the STO crystal. At the lowest approximation, one can ignore the direct coupling between the narrow conducting paths, which are regarded as QWs; see Fig.~\ref{Paper::fig:1} (c). As the resistance occurs dominantly at the tunnel junction between the top FM and the QW, one can ignore the resistance along the QWs and the MTJ structures in Fig.~\ref{Paper::fig:1} (a) and (c) are essentially the same.

\paragraph{Model.}
The MTJ is described by the Hamiltonian
\begin{equation}
\label{Paper::eq:18}
H = \frac{p_x^2+p_z^2}{2m}
+ U(z) - \frac{\alpha(z)}{\hbar}p_x\sigma_y
- \bfDelta(z)\cdot\bf\bfsigma\; ,
\end{equation}
where $\sigma_x$, $\sigma_y$, and $\sigma_z$ are the Pauli matrices.
We have chosen the $x$-axis along the QW axis and the $z$ axis perpendicular to the junction interface [Fig.~\ref{Paper::fig:1} (a)]. The direction of the effective field (``Rashba field'') due to the Rashba SOC is along the $y$ axis. The Rashba SOC is present only on the QW ($0<z<d$):
\begin{equation}
\alpha(z) =
\begin{cases}
\alpha_0 & (0<z<d) \\
0 & \text{(otherwise)}
\end{cases}
\end{equation}
where $d\sim1\unit{nm}$ represents the diameter of the QW or the thickness of the LAO/STO interface. The Zeeman field $\bfDelta(z)$ is due to the ferromagnetism on the top electrode and the QW and is modeled as a vector in the $xy$ plane
\begin{equation}
\label{Paper::eq:13}
\bfDelta(z) =
\begin{cases}
\Delta_1(-\sin\phi,\cos\phi,0) & (z>d)\; , \\
\Delta_2(-\sin\phi,\cos\phi,0) & (0<z<d)\; , \\
0 & (z<0),
\end{cases}
\end{equation}
where the angle $\phi$ ($0<\phi<\pi$) is measured from the $y$-axis (Rashba
field direction).  We assume that $\Delta_1>0$ and that $\Delta_2>0$
and $\Delta_2<0$ for the parallel (P) and anti-parallel (AP) configuration of
the magnetic polarization directions, respectively.
The chemical
potentials (carrier densities) in different regions are described by potential steps and the thin insulating barriers by $\delta$-potentials, giving the potential profile $U(z)$ of the form
\begin{multline}
U(z)
= U_1\theta(z-d) + U_2[\Theta(z-d)-\Theta(z)] \\{}
+ a_bU_b\delta(z-d) + a_b'U_b'\delta(z) \,.
\end{multline}
$U_b$ is responsible for the insulating layer of LAO, $a_b$ is the effective width of the barrier ($a_b\sim 1\text{--}5\unit{nm}$), $U_b'$ is responsible for the junction between the QW and the normal electrode and $a_b'$ is its effective length scale.
For a typical LAO/STO interface \cite{Michaeli12a,Pentcheva09a,Pentcheva07a,Pentcheva08a}, the Fermi energy $E_F\sim 40\unit{meV}$, $\alpha_0\sim \hbar{v_F^0}/8$ with $v_F^0\equiv\sqrt{2E_F/m}$, $\Delta_{2}\sim E_F/16$, and $d\sim 1\unit{nm}$.

The momentum in the $x$-direction is preserved over a tunneling process \cite{endnote:1}. We thus seek a wave function of the form
\begin{math}
\Psi(x,z) = e^{iqx}\psi(z)
\end{math},
where $\psi(z)$ satisfies the 1D Schr\"odinger equation
\begin{math}
H_z\psi(z) = \left(E - {\hbar^2q^2}/{2m}\right)\psi(z)
\end{math}.
The 1D effective Hamiltonian $H_z$ is given by
\begin{equation}
\label{Paper::eq:10}
H_z =
\begin{bmatrix}
1 & 0 \\
0 & 1
\end{bmatrix}
\left(-\frac{\hbar^2}{2m}\frac{d^2}{dz^2} + U_1\right)
- \Delta_1
\begin{bmatrix}
1 & 0 \\
0 & -1
\end{bmatrix}
\end{equation}
in the region $z>d$, by
\begin{multline}
\label{Paper::eq:11}
H_z =
\begin{bmatrix}
1 & 0 \\
0 & 1
\end{bmatrix}
\left(-\frac{\hbar^2}{2m}\frac{d^2}{dz^2} + U_2\right) \\{} -
\begin{bmatrix}
\alpha_0q\cos\phi+\Delta_2 & -i\alpha_0q\sin\phi \\
i\alpha_0q\sin\phi & -(\alpha_0q\cos\phi+\Delta_2)
\end{bmatrix}
\end{multline}
in the region $0<z<d$, and by
\begin{equation}
\label{Paper::eq:12}
H_z =
\begin{bmatrix}
1 & 0 \\
0 & 1
\end{bmatrix}
\left(-\frac{\hbar^2}{2m}\frac{d^2}{dz^2}\right)
\end{equation}
in the region $z<0$.
Here the spin part of $H_z$ has been represented in the eigenbasis $\{\ket{\chi_\up},\ket{\chi_\down}\}$ of $\sigma_y\cos\phi - \sigma_x\sin\phi$ corresponding to the Zeeman field  of the top FM (region $z>d$).
In the region $z>d$, the plane waves of the form
\begin{equation}
\ket{\chi_{\up/\down}}e^{ik_{\up/\down} z} \,,\quad
\ket{\chi_{\up/\down}}e^{-ik_{\up/\down} z}
\end{equation}
with $k_{\up/\down}\equiv\sqrt{2m(E-U_1\pm\Delta_1)/\hbar^2-q^2}$ compose the wave function $\psi(z)$. In the region $0<z<d$, $\psi(z)$ is a linear combination of the plane waves of the form
\begin{equation}
\ket{\chi_\pm}e^{ik_\pm z} \,,\quad
\ket{\chi_\pm}e^{-ik_\pm z}
\end{equation}
where $k_\pm\equiv\sqrt{2m(E-U_2\pm\Delta_2)/\hbar^2-q^2}$ and
\begin{subequations}
\label{Paper::eq:14}
\begin{align}
\ket{\chi_+}
&= \cos(\theta/2)\ket{\chi_\up} + i\sin(\theta/2)\ket{\chi_\down} \\
\ket{\chi_-}
&= i\sin(\theta/2)\ket{\chi_\up} + \cos(\theta/2)\ket{\chi_\down}.
\end{align}
\end{subequations}
Here the angle $\theta$ ($0<\theta<\pi$) switches between $\theta_\mathrm{P}$ and $\theta_\mathrm{AP}$ upon the P ($\theta=\theta_\mathrm{P}$) and AP ($\theta=\theta_\mathrm{AP}$) configuration, which are defined by
\begin{equation}
\tan\theta_\mathrm{P/AP} = \frac{\alpha_0q\sin\phi}{\alpha_0q\cos\phi\pm\Delta_2} \,.
\end{equation}
Imposing proper matching conditions over $\delta$-potentials at $z=0$ and $d$, we determine (both with numerically exact method and with analytically approximate method) the scattering wave function $\psi(z)$ and calculate the TMR ratio,
\begin{math}
\text{TMR} \equiv 1 - {R_\mathrm{P}}/{R_\mathrm{AP}},
\end{math}
where $R_\mathrm{P/AP}$ is the resistance for the P/AP polarization.

\paragraph{Exact Results.}

\begin{figure}[t]
\centering
\includegraphics[width=8.5cm]{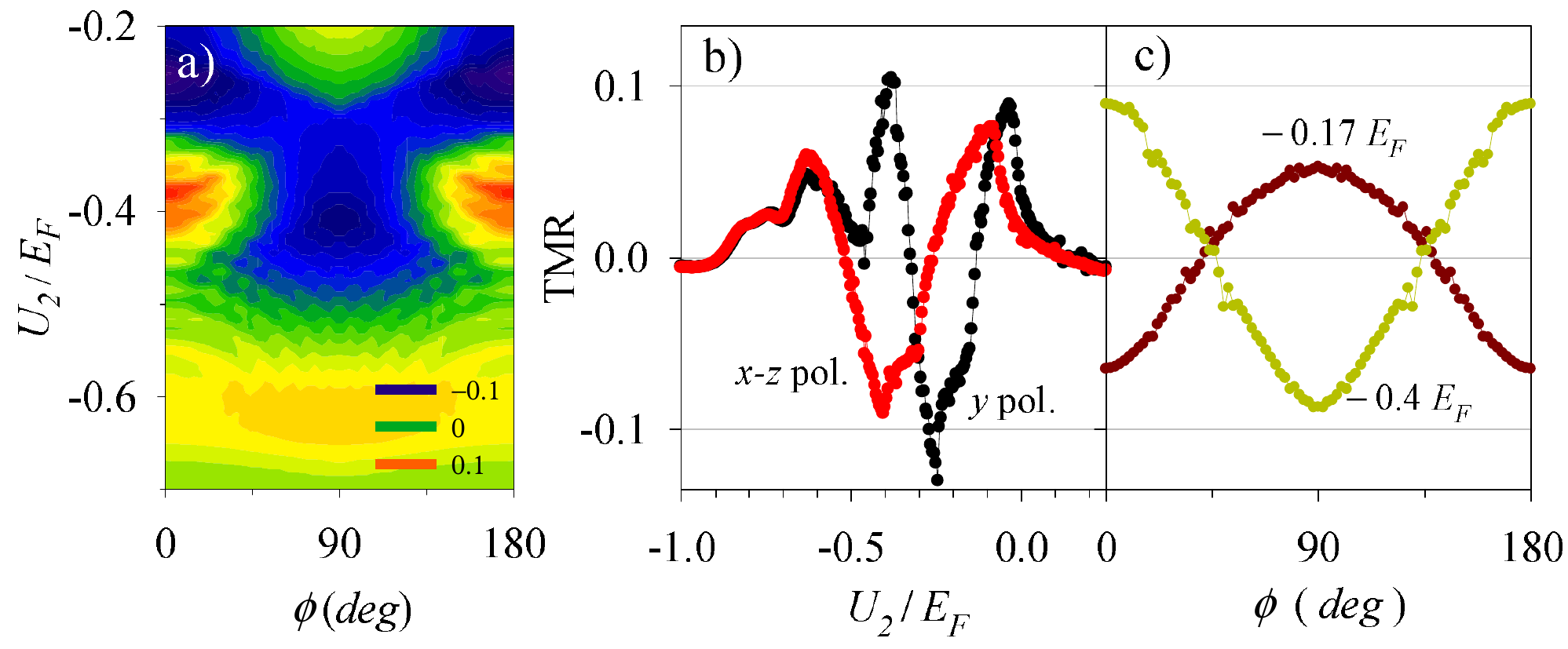}
\caption{(color online) (a) Numerical results of the TMR as a function of $U_2$ and $\phi$ for $d = 4.5/k_F^0$ ($k_F^0\equiv\sqrt{2mE_F/\hbar^2}$), $\alpha_0=\hbar{v_F^0}/8\sqrt{2}$, $\Delta_1=E_F/8$, $\Delta_2=E_F/16$ and $U_1=E_F/4$. (b) Cuts along $\phi=0$ (black dots, $y$ polarization) and $\phi=\pi/2$ (red-gray dots, $x-z$ polarization). (c) Cuts along $\phi$ for the indicated fixed values of $U_2=-0.17 E_F$ and $U_2=-0.40 E_F$.}
\label{Paper::fig:2}
\end{figure}

Figure~\ref{Paper::fig:2} shows the numerically exact results of the TMR as a function of $U_2$ and $\phi$ for a typical set of parameters consistent with the LAO/STO interface \cite{Michaeli12a,Pentcheva09a,Pentcheva07a,Pentcheva08a}.
The numerical method involves the integration of the transmission probabilities over the angle of the incident wave. The algorithm has been devised in such a way to ensure a  sufficient precision for the angular integrals. The details of the numerical method are described in a previous work by one of the authors \cite{Gelabert11a}.

It is shown in Fig.~\ref{Paper::fig:2} that the TMR can be negative as much as $-10\%$. Further, it reveals two additional interesting features: First, the TMR depends rather strongly on $U_2$ [Fig.~\ref{Paper::fig:2} (b)]. Experimentally, $U_2$ corresponds to the backgate voltage and controls the carrier density on the QW (or the LAO/STO interface). In the recent experiment \cite{Ngo14z}, on the other hand, the TMR did not depend much on the gate voltage. However, the actual gate capacitance was not known and it is not clear how large is the actual energy range covered by the gate voltage variation. The gate voltage dependence needs to be tested further. Moreover, in real samples (even if there are twin boundaries) the electric conduction is not completely confined to the narrow paths.

A second remarkable thing 
of Fig.\ \ref{Paper::fig:2}
is the change of the $\phi$ dependence from a $\cos(\phi/2)$ to a $-\cos(\phi/2)$ behavior by tuning the value of $U_2$ [Fig.~\ref{Paper::fig:2} (c)]. This is seen as a reversed change of sign of the TMR when going from $\phi=0$ to $\phi=\pi/2$; from positive to negative for $U_2=-0.4 E_F$, and reversed for $U_2=-0.17 E_F$.

As we discuss below, both features of the exact results 
can be understood qualitatively by means of an analytical (but approximate) method.

\paragraph{Qualitative Features of Single-Barrier Tunneling.}

We first examine the transmission over the \emph{first barrier} at $z=d$.
Before going further, recall the transmission problem of a \emph{spinless particle} with energy $E$ over a potential barrier $U_b$,
\begin{math}
U(z) = U_1\Theta(-z)+U_2\Theta(z) + a_bU_b\delta(z)
\end{math}.
The transmission amplitude $t$ is given by
\begin{equation}
\label{Paper::eq:20}
t(q_b;k_1,k_2) = \frac{\sqrt{k_1k_2}}{(k_1+k_2)/2+iq_b}\; ,
\end{equation}
where
\begin{math}
k_j \equiv \sqrt{{2m(E-U_j)}/{\hbar^2}}
\end{math}
and
\begin{math}
q_b \equiv {ma_bU_b}/{\hbar^2}.
\end{math}
When the barrier is sufficiently high ($U_b\gg E$),
it can be approximated as
\begin{equation}
\label{Paper::eq:2}
t(q_b;k_1,k_2) \approx \frac{\sqrt{k_1k_2}}{iq_b}\; .
\end{equation}
Consider now a scattering state $\psi_\pm(z)$ of the form
\begin{equation}
\psi_\pm(z) =
\begin{cases}
\displaystyle\sum_{s=\up,\down}
\left(A_s\ket{\chi_s}e^{-ik_sz}
  + B_s\ket{\chi_s}e^{ik_sz}\right) & (z>d) \\[5mm]
C_\pm\ket{\chi_\pm}e^{-ik_\pm z} & (z<d)
\end{cases}
\end{equation}
Here we have imposed a boundary condition such that in the region $z<d$ there is only one propagating spin channel $\ket{\chi_\mu}$ of fixed $\mu=\pm$.
On the one hand, the coefficients $A_s$ and $C_\mu$ are related through the transmission coefficients $t_{\mu s}$ by
\begin{math}
C_\mu = \sum_st_{\mu s}A_s
\end{math}.
On the other hand, the matching conditions over the $\delta$-barrier are equivalent to those on the wave function of the form
\begin{align}
\eta_{\mu s}(z) =
\begin{cases}
A_se^{-ik_sz} + B_se^{ik_sz} & (z>d) \\
C_\mu\braket{\chi_s|\chi_\mu}e^{-ik_\mu z} & (z<d)
\end{cases}
\end{align}
imposed separately for each component $s=\up,\down$. This implies by Eq.~(\ref{Paper::eq:20}) that
\begin{math}
C_\mu\braket{\chi_s|\chi_\mu} = A_st(q_b;k_\mu,k_s)
\end{math}.
Combining these two relations leads to
\begin{equation}
\label{Paper::eq:8}
\begin{bmatrix}
1 & 0 \\
0 & 1
\end{bmatrix} =
\begin{bmatrix}
t_{+\up} & t_{+\down} \\
t_{-\up} & t_{-\down}
\end{bmatrix}
\begin{bmatrix}
\frac{\braket{\chi_\up|\chi_+}}{t(q_b;k_+,k_\up)} &
\frac{\braket{\chi_\up|\chi_-}}{t(q_b;k_-,k_\up)} \\
\frac{\braket{\chi_\down|\chi_+}}{t(q_b;k_+,k_\down)} &
\frac{\braket{\chi_\down|\chi_-}}{t(q_b;k_-,k_\down)}
\end{bmatrix}
\end{equation}
Using the approximation~(\ref{Paper::eq:2}),
the matrix on the right hand side of (\ref{Paper::eq:8}) is factorized as
\begin{multline}
\begin{bmatrix}
t_{+\up} & t_{+\down} \\
t_{-\up} & t_{-\down}
\end{bmatrix} \approx i
\begin{bmatrix}
\sqrt{k_+/q_b} & 0 \\
0 & \sqrt{k_-/q_b}
\end{bmatrix} \\{}\times
\begin{bmatrix}
\braket{\chi_+|\chi_\up} &
\braket{\chi_+|\chi_\down} \\
\braket{\chi_-|\chi_\up} &
\braket{\chi_-|\chi_\down}
\end{bmatrix}
\begin{bmatrix}
\sqrt{k_\up/q_b} & 0 \\
0 & \sqrt{k_\down/q_b}
\end{bmatrix}
\end{multline}
The transmission probabilities $T_\mu(q)\equiv\sum_{s}\left|t_{\mu s}\right|^2$ for the channels $\mu=\pm$ are given by
\begin{equation}
\label{Paper::eq:3}
T_\pm(q)
\approx \frac{k_\up+k_\down}{2q_b}\left[
  \frac{k_\pm}{q_b}
  \pm \frac{4m\Delta_1}{\hbar^2(k_\up+k_\down)^2}\cos\theta\right]
\end{equation}
where the $q$-dependence of $k_{\up/\down}$, $k_\pm$ and $\theta$ is implied.
The expressions~(\ref{Paper::eq:3}) for the transmission probabilities between
a ferromagnet and another ferromagnet with strong Rashba SOC is one of our
main results.

\paragraph{Qualitative Features of the Double-Barrier Structure.}

Now we investigate the full double-barrier structure for all possible values of $q$.
For high tunnel barriers, the wave number $k_\pm$ in the central region ($0<z<d$) is quantized to $k_{n}\approx n\pi/d$ ($n=1,2,\cdots$) and the wave function takes the form
\begin{math}
\Psi(x,z) = \ket{\chi_\pm(q_{n,\pm}^\nu)}\sin(k_{n}z)e^{iq_{n,\pm}^\nu x}.
\end{math}
For each $k_{n}$ and a given energy $E$, the allowed values $q_{n,\pm}^\nu$ ($\nu=\lessgtr$) is determined by the dispersion relation
\begin{multline}
E = \frac{\hbar^2}{2m}\left[k_{n}^2 + (q_\pm^\nu)^2\right] + U_2 \\{}
\mp \sqrt{\left(\alpha_0q_{n,\pm}^\nu\right)^2
  + 2\left(\alpha_0q_{n,\pm}^\nu\right)\Delta_2\cos\phi
  +\Delta_2^2}\,.
\end{multline}
Due to narrow confinement ($d\sim 1\unit{nm}$) and strong SOC ($\alpha{q}\simeq E/8$), typically only one $k_n^\pm$ is allowed for each $\pm$. Hereafter we thus drop the subscript $n$: $k\equiv k_{n}$, $q_\pm^\nu\equiv q_{n,\pm}^\nu$ and $\ket{\chi_\pm^\nu}\equiv\ket{\chi_\pm(q_{n,\pm}^\nu)}$. The total transmission probability is given by
\begin{math}
T = \sum_{\mu=\pm}
\left[T_\mu(q_\mu^>)+T_\mu(q_\mu^<)\right].
\end{math}

\begin{figure}
\centering
\includegraphics[width=80mm]{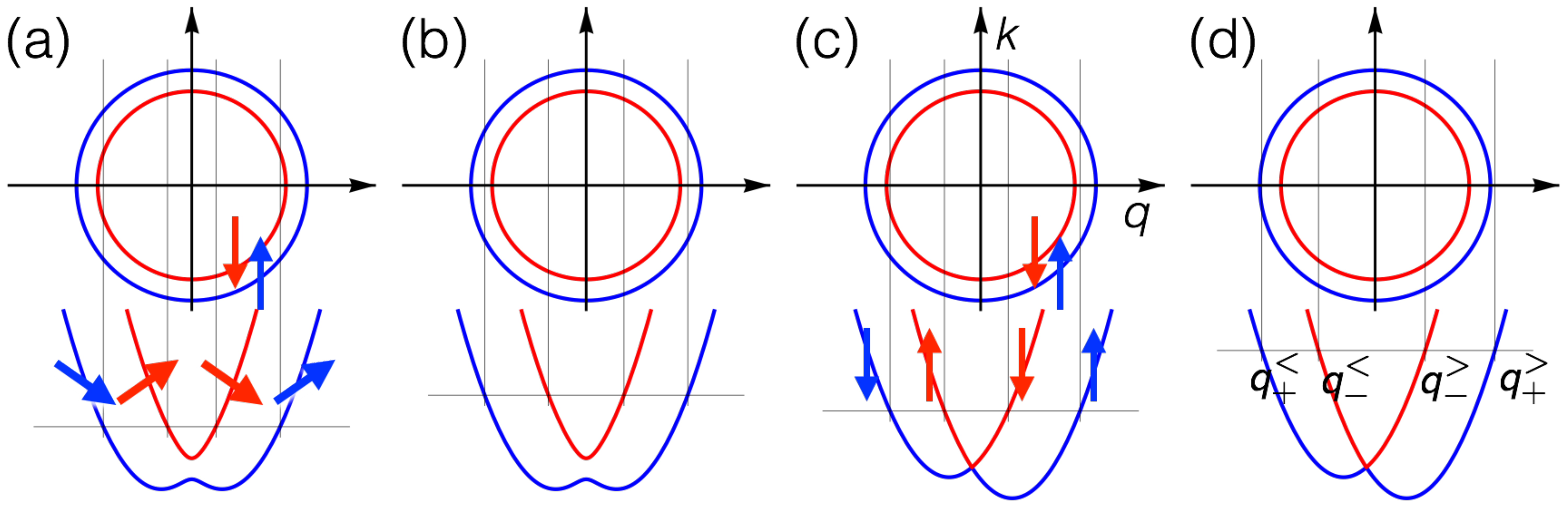}
\caption{(color online) The spin-split Fermi circles in the FM electrode (top) and the dispersion relation in the QW (bottom) for $\phi=\pi/2$ [(a) and (b)] and for $\phi=0$ [(c) and (d)]. The thin horizontal lines indicate the Fermi levels \emph{relative to the band bottoms} and the short arrows depict the spin quantization directions. In (a) and (c) all transverse modes at the Fermi level on the QW contribute to the transport whereas the two outer modes in (b) and the $q_+^>$-mode in (d) do not. }
\label{Paper::fig:3}
\end{figure}

For $\phi=\pi/2$, the Zeeman field is perpendicular to the Rashba field and the dispersion relation is particularly simple. Especially, one has
\begin{math}
q_\pm^> = -q_\pm^< \equiv q_\pm
\end{math},
\begin{math}
q_+ > q_-
\end{math},
\begin{math}
\cos\theta_\mathrm{P}(q_\pm) > 0
\end{math},
\begin{math}
\cos\theta_\mathrm{AP}(q_\pm) < 0
\end{math}, and
\begin{math}
T = 2\left[T_+(q_+)+T_-(q_-)\right]
\end{math}.
When both $q_\pm$ contribute to the transport [Fig.~\ref{Paper::fig:3} (a)],
\begin{multline}
\text{TMR} \propto
\frac{\cos\theta_\mathrm{P}(q_+)-\cos\theta_\mathrm{AP}(q_+)}{k_\up(q_+)+k_\down(q_+)} \\{}
- \frac{\cos\theta_\mathrm{P}(q_-)-\cos\theta_\mathrm{AP}(q_-)}{k_\up(q_-)+k_\down(q_-)}
\end{multline}
The $q_+$ ($q_-$) channel contributes a positive (negative) TMR. As $q_+>q_-$, $k_\up(q_+)+k_\down(q_+)<k_\up(q_-)+k_\down(q_-)$ and the positive contribution from $q_+$-channel dominates.
When $q_+$-channel is not allowed [Fig.~\ref{Paper::fig:3} (b)], $T_-(q_-)$ from the $q_-$ channel is the sole contribution and the TMR becomes negative.

For $\phi=0$ ($\phi=\pi$), $\theta_\mathrm{P}(q_\pm^\gtrless)=\theta_\mathrm{AP}(q_\pm^\gtrless)=0$ and the total transmission reads as [Eq.~(\ref{Paper::eq:3}) with $k_\pm=k$]
\begin{equation}
T = \frac{k}{q_b^2}\left[
  k_\down(q_-^>) + k_\up(q_-^<)
  + k_\down(q_+^<) + k_\up(q_+^>) \right].
\end{equation}
Note that
\begin{math}
q_+^> > -q_+^< > -q_-^< > q_-^> > 0
\end{math}
in the P polarization configuration [Fig.~\ref{Paper::fig:3} (c) and (d)].
The TMR is then given by
\begin{multline}
\label{Paper::eq:17}
\text{TMR} \propto
  [k_\up(q_+^>)-k_\down(q_+^>)]
  -[k_\up(q_+^<) - k_\down(q_+^<)] \\{}
  + [k_\up(q_-^<)-k_\down(q_-^<)]
  -[k_\up(q_-^>) - k_\down(q_-^>)] 
\end{multline}
where the terms have been arranged in decreasing order (all values within square brackets are positive) and all $q_\pm^\gtrless$ have been defined for the P polarization configuration.
As $U_2$ (the chemical potential in the central region) varies, the $q_+^>$ channel may become disallowed [Fig.~\ref{Paper::fig:3} (d)]. In such a case, there are more negative contributions to the TMR. As $U_2$ varies further, the $q_+^<$ channel is also disallowed, and the TMR becomes positive again. As $U_2$ varies even further, the $q_-^<$ channel stops contributing to the transport and the TMR becomes negative once more.

Putting all together, with $U_2\to-\infty$, TMR is positive both at $\phi=0$ and $\phi=\pi/2$. As $U_2$ moves up, the $q_+^>$-mode at $\phi=\pi/2$ gets disallowed first at $U_2\approx-0.6E_F$; 
the TMR($\phi=\pi/2$) becomes negative but TMR(0) remains positive. At $U_2\approx-0.5E_F$, the $q_+^>$ mode at $\phi=0,\pi$ gets disallowed and both TMR$(\pi/2)$ and TMR$(0)$ become negative. But quite soon at $U_2\approx-0.45E_F$, the $q_+^<$ mode gets disallowed and TMR(0) quickly becomes positive again. Therefore, until $U_2\approx -0.2E_F$, where both spin channels get disallowed, TMR($\pi/2$) and TMR(0) remain negative and positive, respectively. As a function of $\phi$, the TMR is expected to behave like $\cos(\phi/2)$. This is consistent with Fig.~\ref{Paper::fig:2} (b) and (c) for $U_2\lesssim -0.2E_F$.

We stress that in these qualitative arguments, evanescent waves have
been ignored completely. In particular, for $U_2\gtrsim-0.2E_F$ (with other
parameters fixed as given), both spin channels are evanescent \cite{endnote:3}
in the central region ($0<z<d$) and cannot be addressed within the approximate analytical method.
Quite interestingly, as we have seen above, the contributions of the evanescent waves are highly nontrivial in this parameter range and give rise to $-\cos(\phi/2)$ behavior.

\paragraph{Conclusion.}

We have considered a double-barrier MTJ consisting of a ferromagnetic electrode, a QW with magnetic ordering and strong Rashba spin-orbit coupling, and a normal metal electrode where the junction is formed on the cylindrical shell of the QW. The structure may have a relevance as a simplified model for the magnetic tunnel junction with a LAO/STO transition metal oxide interface including twin boundaries. The latter has been reported to exhibit sign-switching anisotropic TMR.
By means of both qualitative analysis and numerically exact calculations, we have shown that our model exhibits a sign-switching anisotropic TMR. The negative TMR occurs as a combined effect of one-dimensionality, magnetic order, and strong SOC in the QW.

\paragraph{Acknowledgments.}
This work was supported by the BK21 Plus Project from the Korean Government and by MINECO (Spain) Grant FIS2011-23526.

\bibliographystyle{apsrev4-1}
\bibliography{Paper}

\begin{thebibliography}{25}
\bibitem[{end({\natexlab{a}})}]{endnote:2}
\bibinfo{note}{The transport is thus across (not along) the axis of the QW.}
\bibitem[{end({\natexlab{b}})}]{endnote:1}
\bibinfo{note}{Here we assume that the junction (QW) is infinitely wide
  (long).}
\bibitem[{end({\natexlab{c}})}]{endnote:3}
\bibinfo{note}{Even if the energy is positive, evanescent waves appear already
  for negative $U_2$ because of the strong Rashba SOC in the centeral region
  ($0<z<d$).}
\end{thebibliography}

\end{document}